\documentclass[12pt,preprint]{aastex}
\usepackage{graphicx}
\shorttitle{Fluid-Kinetic Scales Turbulence Link}
\shortauthors{Telloni \& Bruno}
\begin{document}
\title{Linking Fluid and Kinetic Scales in Solar Wind Turbulence}
\author{D. Telloni\altaffilmark{1} and R. Bruno\altaffilmark{2}}
\altaffiltext{1}{National Institute for Astrophysics, Astrophysical Observatory of Torino, Via Osservatorio 20, 10025 Pino Torinese, Italy}
\altaffiltext{2}{National Institute for Astrophysics, Institute for Space Astrophysics and Planetology, Via del Fosso del Cavaliere 100, 00133 Roma, Italy}
\begin{abstract}
We investigate possible links between the large-scale and small-scale features of solar wind fluctuations across the frequency break separating fluid and kinetic regimes. The aim is to correlate the magnetic field fluctuations polarization at dissipative scales with the particular state of turbulence within the inertial range of fluctuations. We found clear correlations between each type of polarization within the kinetic regime and fluid parameters within the inertial range. Moreover, for the first time in literature, we showed that left-handed and right-handed polarized fluctuations occupy different areas of the plasma instabilities-temperature anisotropy plot, as expected for Alfv$\acute{\textrm{e}}$n Ion Cyclotron and Kinetic Alfv$\acute{\textrm{e}}$n waves, respectively.
\end{abstract}
\section{Introduction}
\label{sec:introduction}
The high frequency break often found in interplanetary magnetic field power density spectra separates the fluid from the kinetic regime of fluctuations. This break is located around scales which are typical of protons kinetics like the proton inertial length $\lambda_{i}=c/\omega_{p}$ and the proton Larmor radius $\lambda_{L}=v_{th}/\Omega_{p}$, where $\omega_{p}$ is the local plasma frequency while $\Omega_{p}$ is the local gyro-frequency, with $v_{th}$ and $c$ the thermal speed and the speed of light, respectively. Several authors tried to reproduce the location of the frequency break according to various models \citep{leamon1998,perri2010,bourouaine2012} but \citet{markovskii2008} showed that none of the available model could predict a value for the frequency break in good agreement with the observations. Only recently, \citet{bruno2014a}, studying the radial behavior of the frequency break location within the expanding high speed wind, were able to conclude that the resonant condition for outward parallel propagating Alfv$\acute{\textrm{e}}$n/ion-cyclotron waves (ICWs) was the mechanism able to provide the best agreement with the observations. This result, although not expected on the basis of anisotropy predictions by any turbulent cascade \citep{chen2014}, gave new relevance to the ion-cyclotron resonance mechanism in the frame of turbulence collisionless heating without canceling the role of other possible mechanisms that might be at work as well. \citet{leamon1999} proposed Landau damping of obliquely propagating kinetic Alfv$\acute{\textrm{e}}$n waves (KAWs) resulting in a frequency break corresponding to the scale of the Larmor radius $\lambda_{L}$; for \citet{dmitruk2004} 2D turbulence dissipation through turbulence reconnection process and generation of current sheets of the order of the ion inertial length $\lambda_{i}$ enhances the role of this scale, which is the most relevant one also in the framework of incompressible Hall MagnetoHydroDynamics (MHD) used by \citet{galtier2006} to explain the break.

In this respect, it is important to investigate the nature of the fluctuations beyond the frequency break. Studies based on magnetic field compressibility suggested the presence of KAWs and/or whistler waves \citep{leamon1998,gary2004,alexandrova2008,hamilton2008,kiyani2009,sahraoui2009,salem2012,tenbarge2012}. Intermittency analyses of magnetic fluctuations between proton and electron scales based on observations \citep{alexandrova2008,perri2012,karimabadi2013} together with a statistical correspondence between intermittent events, identified as current sheets generated by turbulence cascade, and cases of enhanced temperature in direct numerical simulations of a hybrid Vlasov-Maxwell model \citep{greco2012,servidio2012} suggested a relevant role of coherent structures for heating processes. On the other hand, several studies on the polarization of magnetic fluctuations beyond the frequency break \citep{goldstein1994,leamon1998,hamilton2008} clearly reported a right-handed polarization state, attributed to KAWs. Later, more sophisticated analyses unraveled the presence of both right- and left-handed fluctuations \citep{he2011,podesta2011,he2012a,he2012b}, generally attributed to KAWs and outward propagating ICWs, respectively, although \citet{podesta2011} did not exclude the possibility of inward propagating whistler modes for the left-handed polarization. Recently, \citet{telloni2015} and \citet{bruno2015} showed that, within a high speed stream, right-handed fluctuations are characterized by higher power level and, in particular, by a higher compressibility with respect to the left-handed counterpart, favoring the ICWs interpretation for the latter ones. Moreover, the same authors showed that, moving from fast to slow wind, within the same stream, the spectral slope of the kinetic regime becomes shallower and the polarization of the fluctuations is depleted with the left-handed modes being the first ones to disappear. The above observations leave many questions still unanswered, especially those related to the origin and the role of these fluctuations with respect to the heating mechanisms at work.

Fast and Alfv$\acute{\textrm{e}}$nic wind shows strong temperature anisotropy in favor of the perpendicular direction which is likely to be caused by the cyclotron resonance with Alfv$\acute{\textrm{e}}$n left-handed waves \citep{hollweg2002,marsch2006}. Wave-particle interaction due to plasma instabilities modify the proton velocity distribution \citep[][and references therein]{hellinger2006}. Plasma instabilities are predicted to develop during the expansion of the wind even if the expansion is not adiabatic, as requested by the CGL model \citep{chew1956}, due to the radial dependence of the wind parameters, which modifies the temperature anisotropy and the parallel proton plasma beta $\beta_{\parallel}$. Consequently, an initially isotropic, low beta plasma, stable with respect to plasma instabilities, would become unstable. \citet{matteini2007} found that the fast wind plasma, which is characterized by a clear anti-correlation between temperature anisotropy and $\beta_{\parallel}$ at $0.3$ AU \citep{marsch2004}, evolves towards the fire-hose instability regions with increasing distance. On the other hand, the same authors did not find any clear evolution for the slow wind. As a matter of fact, we have a rather complete characterization regarding both plasma and magnetic field parameters and fluctuations only at scales larger than the proton scales, i.e. within the inertial range. Moreover, it is well known that, within this range of scales, fast wind largely differs from slow wind especially for the nature of the fluctuations, which are less compressive and more Alfv$\acute{\textrm{e}}$nic. In this paper we are devoted to investigate any possible link between the large-scale and small-scale features of solar wind fluctuations across the high frequency break. In particular, we aim to verify whether in correspondence of highly Alfv$\acute{\textrm{e}}$nic fluctuations within the inertial range, we find any particular state of plasma within the $\beta_{\parallel}$-temperature anisotropy diagram and any particular polarization of magnetic field fluctuations at kinetic scales.

In order to accomplish this task we will exploit the possibility to single out events characterized by different polarization of magnetic field fluctuations for a selected interval within a high speed stream observed at $1$ AU \citep{bruno2015}.
\section{Data analysis and results}
\label{sec:data_analysis_results}
\begin{figure*}
	\centering
	\includegraphics[width=\textwidth]{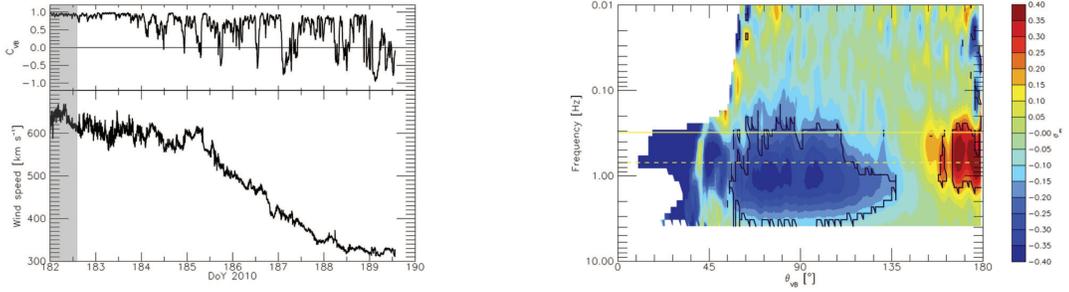}
	\caption{Correlation coefficient $C_{VB}={\sigma_{c}}/\sqrt{1-\sigma_{r}^{2}}$ evaluated within a sliding window of $1\,hr$ (\emph{upper-left}). Speed profile of the fast wind stream (\emph{lower-left}). The opaque area indicates the analyzed time interval. 2D histogram of the reduced normalized magnetic helicity versus frequency and sampling angle with respect to the local direction of the background magnetic field (\emph{right}). The frequency locations corresponding to the frequency break (solid yellow line) and the selected frequency of $0.7\,Hz$ (dashed yellow line) are also shown.}
	\label{fig:time_interval}
\end{figure*}

The selected time interval on which we performed our analysis spans over the first $13.4$ hours of DoY 182 (July $1^{\textrm{st}}$) of 2010, corresponding to $2^{19}$ magnetic field data points with a cadence of $92\,ms$ recorded by the magnetic field experiment \citep{lepping1995} aboard the WIND s/c. This time interval is highlighted by the opaque area in the lower-left panel of Fig. \ref{fig:time_interval} which shows $92\,s$ averages of plasma speed within the fast stream as observed by the solar wind plasma experiment \citep{ogilvie1995}. This stream, already considered in previous papers \citep{bruno2014a,bruno2014b,telloni2015,bruno2015}, is a typical corotating fast stream characterized by high amplitude Alfv$\acute{\textrm{e}}$nic fluctuations within its trailing edge and less Alfv$\acute{\textrm{e}}$nic fluctuations within its rarefaction region. This is shown by the upper-left panel of Fig. \ref{fig:time_interval}, which reports the value of the correlation coefficient $C_{VB}={\sigma_{c}}/\sqrt{1-\sigma_{r}^{2}}$ between velocity and magnetic field fluctuations within a sliding window of $1\,hr$ \citep{bavassano1998}, with $\sigma_{r}\neq\pm1$. The normalized cross-helicity $\sigma_{c}$ and residual energy $\sigma_{r}$ are defined as $H_{c}/E$ and $E_{r}/E$, respectively, being $H_{c}=<\mathbf{v}\cdot\mathbf{b}>$ the cross-helicity, $E_{r}=<\mathbf{v}^{2}-\mathbf{b}^{2}>$ the residual energy and $E=\frac{1}{2}<\mathbf{v}^{2}+\mathbf{b}^{2}>$ the total energy, where $\mathbf{v}$ and $\mathbf{b}$ are the fluctuations of velocity and magnetic field, respectively, this last one expressed in Alfv$\acute{\textrm{e}}$n units ($\mathbf{b}\longrightarrow\frac{\mathbf{b}}{\sqrt{4\pi\rho}}$) being $\rho$ the mass density, and the angle brackets indicate an ensemble average. Alfv$\acute{\textrm{e}}$nic fluctuations are expected to show values of $|\sigma_{c}|\sim1$ and $\sigma_{r}\sim0$, which denote magnetic and velocity fluctuations alignment and magnetic and kinetic energy equipartition, respectively. The normalized cross-helicity has been multiplied by $-1$ so that a positive value would indicate an outward sense of propagation.

The right panel shows a 2D histogram of the reduced normalized magnetic helicity versus frequency and sampling angle with respect to the local direction of the background magnetic field. Details on the analysis are largely reported in previous papers \citep{telloni2015,bruno2015} and are similar to the numerical techniques used by other authors like \citet{bruno2008}, \citet{horbury2008}, \citet{he2011} and \citet{podesta2011}. The black contours indicate the $95$\% confidence interval. The blue and red populations, located roughly between $0.3$ and $4\,Hz$, correspond to right and left-handed polarized fluctuations and have been recognized as due to KAWs and ICWs, respectively \citep{he2011,podesta2011,klein2014,telloni2015,bruno2015}. The solid yellow line shows the frequency location of the break of the magnetic power spectrum at $0.31\,Hz$. For this particular time interval, the ion-cyclotron resonance frequency and the frequencies corresponding to the ion-inertial length and Larmor radius are $0.29\,Hz$, $0.61\,Hz$ and $0.55\,Hz$, respectively. As already shown in \citet{bruno2014b}, the ion-cyclotron resonance frequency is the one that better matches the location of the break point.

\begin{figure}
	\centering
	\includegraphics[width=0.9\hsize]{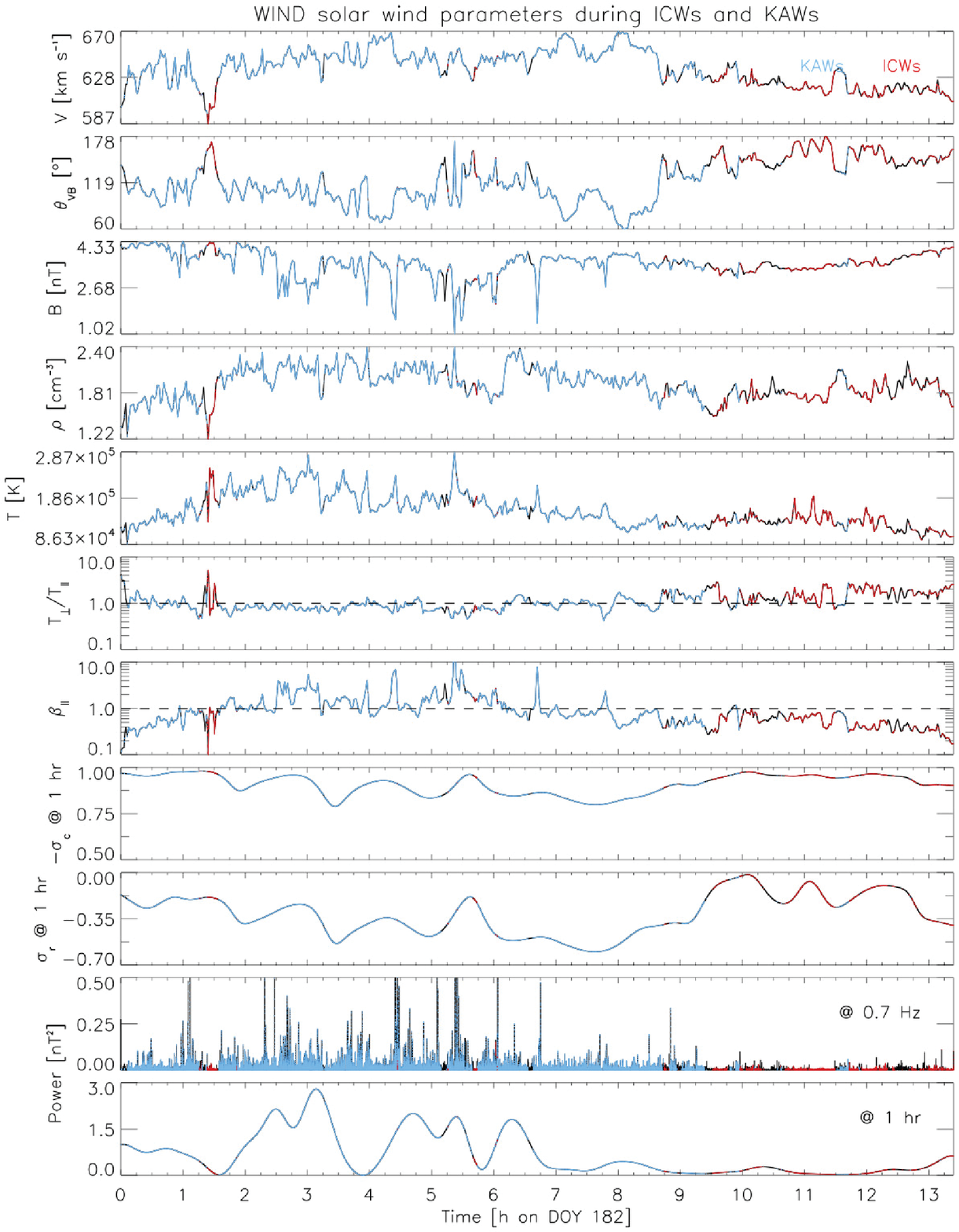}
	\caption{\emph{From top to bottom} Wind speed, sampling angle, magnetic field strength, proton number density, total proton temperature, temperature anisotropy $T_{\perp}/T_{\parallel}$, proton plasma $\beta_{\parallel}$ at the scale of $92\,s$, $-\sigma_{c}$ and $\sigma_{r}$ at the scale of $1\,hr$, power of the magnetic field strength fluctuations at $0.7\,Hz$ and at $1\,hr$. KAW and ICW events are marked in blue and red colors, respectively.}
	\label{fig:icws_kaws}
\end{figure}

Since our numerical technique allows us to locate and select any event in both time and frequency/scale domains, we decided to select a particular frequency and extract all the events corresponding to KAWs and ICWs, in order to highlight any possible correlation with macroscopic plasma and magnetic field parameters (Fig. \ref{fig:icws_kaws}). The dashed yellow line in the right panel of Fig. \ref{fig:time_interval} indicates the selected frequency of $0.7\,Hz$. This is an arbitrary choice, but it allows us to sample a large number of KAWs and ICWs since this frequency runs close to the cores of the two populations.

Data shown in Fig. \ref{fig:icws_kaws} are $92\,s$ averages of the merged plasma and magnetic field parameters. From top to bottom, we show wind speed, sampling angle, magnetic field strength, proton number density, total temperature $T=(2T_{\perp}+T_{\parallel})/3$, temperature anisotropy $T_{\perp}/T_{\parallel}$, proton plasma $\beta_{\parallel}=v_{th_{\parallel}}^{2}/v_{A}^{2}$ ($v_{th_{\parallel}}$ and $v_{A}$ are the parallel thermal speed and the Alfv$\acute{\textrm{e}}$n speed, respectively), normalized cross-helicity $-\sigma_{c}$ and residual energy $\sigma_{r}$ evaluated within a sliding window of $1\,hr$, and power of the magnetic field strength fluctuations at $0.7\,Hz$ and at the scale of $1\,hr$. The clear anticorrelation between speed and $\theta_{VB}$ highlights that observed turbulence features depend on the sampling direction \citep{matteini2014}. Moving along the field would help to capture fluctuations with $\kappa_{\parallel}$ rather than fluctuations with $\kappa_{\perp}$ although their presence, in spite of the sampling direction, is regulated by other factors, e.g. plasma beta and temperature anisotropy. As a matter of fact, the first $3/4$ of this time interval are generally characterized by larger fluctuations in wind speed, magnetic field strength and density, temperature anisotropy $<1$, higher total temperature, $\beta_{\parallel}>1$, less Alfv$\acute{\textrm{e}}$nic fluctuations dominated by magnetic energy at fluid scales, and higher compressive magnetic fluctuations at both kinetic and fluid scales. On the opposite, the last quarter of this interval has much smaller fluctuations in the wind parameters, temperature anisotropy $>1$, $\beta_{\parallel}<1$, and more Alfv$\acute{\textrm{e}}$nic and less compressive fluctuations at fluid and kinetic scales.

Most of the KAW events (blue color) populate the first $3/4$ of the interval while most of the ICW events (red color) reside within the last quarter. Interesting to notice is also the presence of a clear ICW event between hour $1$ and $2$, immersed in a sea of KAWs, just where the temperature anisotropy $>>1$ and the $\beta_{\parallel}<<1$, in agreement with expectations.

\begin{figure}
	\centering
	\includegraphics[width=\hsize]{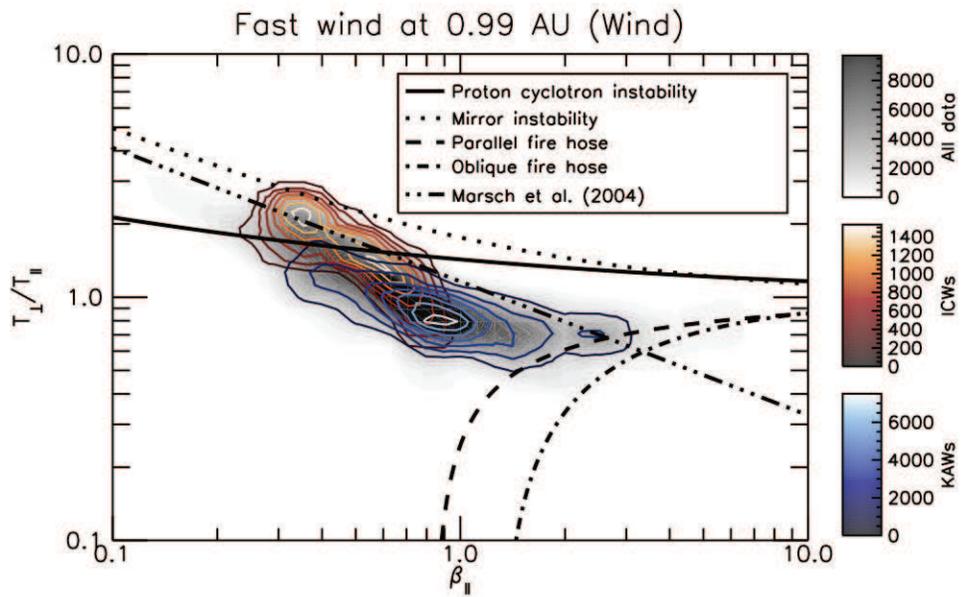}
	\caption{Proton temperature anisotropy vs proton $\beta_{\parallel}$ histogram. Blue and red contours highlight KAW and ICW events, respectively. The borders where various plasma instabilities should limit the proton distribution are marked with different line types, as directly listed within the figure.}
	\label{fig:temperature_anisotropy_vs_parallel_beta}
\end{figure}

These results allow us to represent this situation in a $\beta_{\parallel}$-temperature anisotropy 2D histogram (Fig. \ref{fig:temperature_anisotropy_vs_parallel_beta}) as firstly reported by \citet{gary2001}, \citet{kasper2002} and \citet{hellinger2006}. Blue and red contours highlight KAW and ICW events, where the color saturation corresponds to the number of cases. The different solid and dashed lines indicate the possible plasma instabilities as adopted from \citet{hellinger2006} for a maximum growth rate $\gamma\sim10^{-3}\omega_{c}$, where $\omega_{c}$ is the local ion cyclotron frequency.

In case of substantial departure from an isotropic, homogeneous, thermal equilibrium, plasma instabilities develop in order to reshape the particle velocity distribution towards a Maxwellian. For $T_{\perp}/T_{\parallel}>1$ cyclotron \citep[e.g.][]{gary1994} and mirror \citep[e.g.][]{pokhotelov2004} instabilities arise. On the other hand, for $T_{\perp}/T_{\parallel}<1$ other instabilities like parallel \citep{quest1996} and oblique \citep{hellinger2000} fire hose develop.

Although the distribution shown in Fig. \ref{fig:temperature_anisotropy_vs_parallel_beta} spans a considerable interval of temperature anisotropy and plasma $\beta_{\parallel}$, the two populations are clearly distinguishable. KAWs are mainly located around $\beta_{\parallel}\sim1$ and $T_{\perp}/T_{\parallel}\sim1$, while ICWs lay along the inverse correlation curve found, for core protons, by \citet{marsch2004} for anisotropy values $>1$ and $\beta_{\parallel}<1$. This relation was found within high speed streams at short heliocentric distances where the Alfv$\acute{\textrm{e}}$nic character of the fluctuations in the inertial range is stronger. However, we are able to identify a strong, although minority, ICW population, obeying the same empirical relationship, even at $1$ AU, where the Alfv$\acute{\textrm{e}}$nicity of the fluctuations is somewhat depleted \citep{bruno1985}. The ICW population lies inside the region unstable with respect to the proton cyclotron instability. This disagreement between the observations and linear kinetic theory may reside in how the proton cyclotron instability is derived. More accurate description of the solar wind plasma composition and of the proton distribution function, and taking properly into account nonlinear effects, may indeed place the threshold at higher values in the temperature parallel beta-anisotropy plane and provide a better agreement with observations \citep[see also][]{hellinger2006,matteini2012,matteini2013}.

Thus, the presence of the fluctuations that we identified as ICWs might be a direct consequence of the presence of well defined Alfv$\acute{\textrm{e}}$nic fluctuations that we found in the inertial range during the last quarter of the time interval.

\begin{figure}
	\centering
	\includegraphics[width=\hsize]{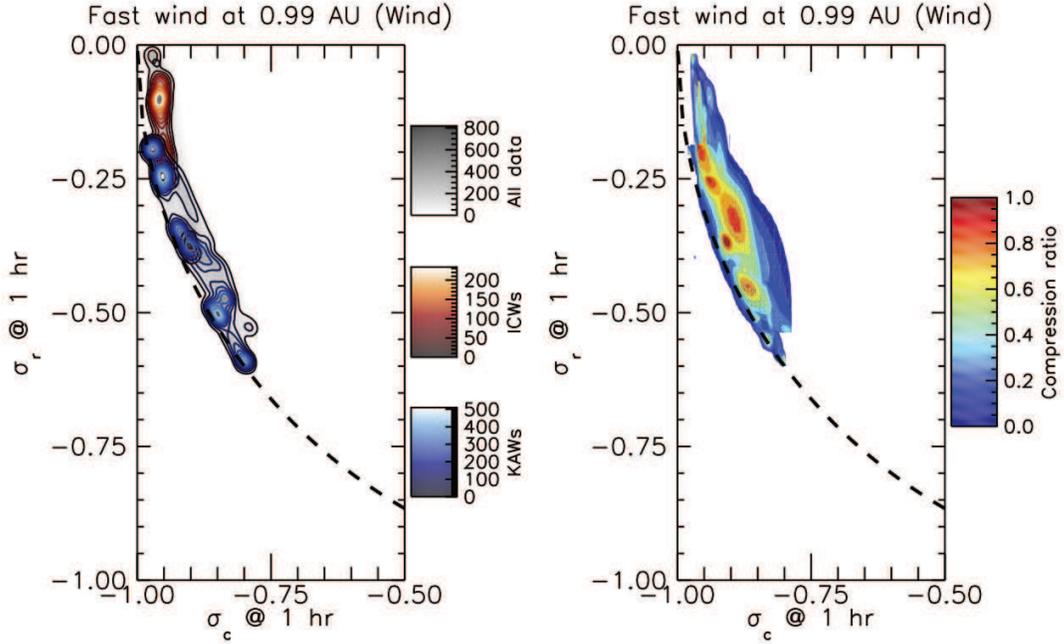}
	\caption{Location in the $1\,hr$ scale $\sigma_{c}-\sigma_{r}$ plane of the magnetic field polarization state (\emph{left}) and the compression ratio (\emph{right}) at $0.7\,Hz$. Blue and red contours indicate, as for Fig. \ref{fig:temperature_anisotropy_vs_parallel_beta}, KAWs and ICWs, respectively. The black dashed curves indicate the limit of validity for the functional relation $\sigma_{c}^{2}+\sigma_{r}^{2}\leq1$.}
	\label{fig:sigma_c_vs_sigma_r}
\end{figure}

This conclusion can be derived also in a different way, i.e. reporting our results in a $\sigma_{c}-\sigma_{r}$ plane \citep{bavassano1998}, associating to right or left-handed polarized events at the frequency of $0.7\,Hz$ the corresponding value of $\sigma_{c}$ and $\sigma_{r}$ at $1\,hr$ scale. In doing so, we can build the 2D histogram of the left panel of Fig. \ref{fig:sigma_c_vs_sigma_r}, where blue and red contours indicate, as for Fig. \ref{fig:temperature_anisotropy_vs_parallel_beta}, KAWs and ICWs, respectively.

Similarly to the results shown in the previous figure, the two populations appear widely separated along the limiting circle, highlighted by the black dashed curve, given by the functional relation $\sigma_{c}^{2}+\sigma_{r}^{2}\leq1$ \citep{bavassano1998}. This means that while ICWs appear during time intervals characterized by rather pure Alfv$\acute{\textrm{e}}$nic turbulence, KAWs populate time intervals characterized by fluctuations that still have a high degree of magnetic and velocity fluctuations alignment, but do not show equipartition being magnetically dominated.

In terms of compressibility at kinetic scales ($0.7\,Hz$), expressed as the ratio between the power associated to strength fluctuations and the total magnetic energy \citep{bavassano1982}, the right panel of Fig. \ref{fig:sigma_c_vs_sigma_r} reveals that, as expected, KAWs are more compressive than ICWs \citep{telloni2015,bruno2015}.
\section{Summary and discussion}
\label{sec:summary_discussion}
One of the most important results in space plasma turbulence would be that of linking fluid and kinetic scales to understand how energy is reversed towards the dissipation range. This Letter represents a promising attempt in this direction. We analyzed a long enough time interval within the trailing edge of a fast stream observed by WIND. We focused on kinetic scales right after the frequency break and analyzed the polarization status of magnetic fluctuations. We found two distinct families of fluctuations, roughly between $0.3$ and $4\,Hz$, characterized by right- and left-handed polarization and interpreted as KAWs and ICWs, respectively \citep[see also][]{telloni2015}.

The detected KAWs and ICWs were mainly associated with time intervals characterized by different values of proton temperature anisotropy and $\beta_{\parallel}$. Thus, for the first time to our knowledge, it was possible to separate, within the $\beta_{\parallel}-T_{\perp}/T_{\parallel}$ plane, magnetic field fluctuations with different polarization status. As expected, we found that these two distinct populations occupy different regions of this 2D plane. Most of the left-handed population, identified as ICWs, show a $\beta_{\parallel}$ smaller than $0.7$ and a temperature anisotropy generally higher than $1$. The core of this population shows an anti-correlation similar to the empirical relation found by \citet{marsch2004} and, as such, is limited neither by the mirror instability nor by the proton cyclotron instability. The right-handed population, identified as KAWs, is mainly located around $\beta_{\parallel}\sim1$ and $T_{\perp}/T_{\parallel}\leq1$ and lies below the proton-cyclotron instability and possibly limited by the parallel/oblique fire hose instability.

We found robust hints that the status of the fluctuations within the kinetic range is somehow related to the status of turbulence within the inertial range. As a matter of fact, ICWs events at kinetic scales were also characterized, at fluid scales, by fluctuations with an enhanced Alfv$\acute{\textrm{e}}$nic character ($|\sigma_{c}|\sim1$ and $|\sigma_{r}|\sim0$). On the other hand, KAWs events were found to be associated with fluid-scale fluctuations that showed the typical Alfv$\acute{\textrm{e}}$nic correlation between magnetic and velocity fluctuations, but dominated by an excess of magnetic energy, being characterized by high values of $|\sigma_{c}|$ but negative values of $\sigma_{r}$. The two populations are characterized at kinetic scales also by a different level of magnetic compressibility, being the KAWs more compressive than the ICWs.

One possibility for the presence of ICWs might be related to the presence of Alfv$\acute{\textrm{e}}$nic fluctuations within the inertial range, which would experience cyclotron resonance with protons \citep{bruno2014a} causing the build up of the observed temperature anisotropy. This anisotropy reshapes the particle velocity distribution function and represents the free energy successively released via the ICWs generation \citep{hellinger2006}.

However, the presence of ICWs is not correlated to a higher total plasma temperature. In our case, higher temperature is observed within the plasma region where KAWs have been identified, which, on the other hand, carry more power than the ICWs, as shown in the previous section and also reported by \citet{telloni2015}. As a matter of fact, when the parallel component of the fluctuations becomes smaller and the perpendicular one dominates, Landau resonance plays the most important role and can exploit the larger power content of these fluctuations \citep{telloni2015}, which can be generated by a non-linear decay of a pump Alfv$\acute{\textrm{e}}$n wave or non-linearly excited by fast magneto-acoustic waves \citep{voitenko2005} or naturally generated by the turbulent cascade as this reaches ion scales \citep[e.g.][]{franci2015}, without the need for a specific wave-coupling mechanism. In other words, when the oblique component is strongly dominant within the inertial range, KAWs play the major role, generating compressive fluctuations in the magnetic field that we observe at the same time at both fluid and kinetic scales.

The results we showed, although relative to a single time interval, are representative of the solar wind state within the trailing edge of this corotating high speed stream, mostly similar to other fast streams observed at $1$ AU. It would then be interesting to follow the evolution of the fluctuations within the $\beta_{\parallel}-T_{\perp}/T_{\parallel}$ plane during the radial expansion of the wind. A similar study \citep{matteini2007}, although without the polarization analysis, showed that at $0.3$ AU the whole population is characterized by a well defined anti-correlation between temperature anisotropy $T_{\perp}/T_{\parallel}$ and $\beta_{\parallel}$, while farther out in the heliosphere the situation changes in favor of a population mainly constrained by the fire hose instability. By comparing these results with ours, we might be able to predict that at short heliocentric distances, within fast wind, fluctuations at proton scales would show an enhanced population of left-handed polarized ICWs. This inference would nicely agree with the enhanced Alfv$\acute{\textrm{e}}$nic character of turbulence expected at shorter heliocentric distances \citep{bruno1985} at fluid scales. This prediction would be easily tested by future space missions such as Solar Orbiter and Solar Probe Plus.
\section*{Acknowledgements}
We acknowledge useful discussions with L. Trenchi. DT is financially supported by the Italian Space Agency (ASI) under contract I/013/12/0. Data from WIND were obtained from the NASA-CDAWeb website.

\end{document}